\documentclass[a4paper,preprint,12pt]{revtex4}

\usepackage[tbtags]{amsmath}
\usepackage{amssymb}
\usepackage{amsthm}
\usepackage{fancyhdr}
\usepackage[dvips]{graphicx}
\usepackage{exscale}
\usepackage[bookmarks]{hyperref}

\allowdisplaybreaks[1]

\begin{document}
\title{Semi-numerical power expansion of Feynman integrals}
\author{Volker Pilipp}
\affiliation{Institute of Theoretical Physics\\ Universit\"at Bern\\
             Sidlerstrasse 5, CH-3012 Bern }
\email{volker.pilipp@itp.unibe.ch}

\begin{abstract}
I present an algorithm based on sector decomposition and Mellin-Barnes
techniques to power expand Feynman integrals. The coefficients of 
this expansion are given in terms of finite integrals that can be
calculated numerically. I show in an example the benefit of this
method for getting the full analytic power expansion from differential
equations by providing the correct ansatz for the solution. For method
of regions the presented algorithm provides a numerical check, which
is independent from any power counting argument. 
\end{abstract}

\maketitle
\section{Introduction}
For power expanding Feynman integrals several methods exist, where all
of them have their limitations. Mellin-Barnes techniques provides a
very general method to obtain all powers
\cite{Greub:1996tg,Smirnov:2002pj}. 
This method however fails if the integrals are
getting too complex. On the other hand \emph{method of regions} 
\cite{Smirnov:2002pj, Gorishnii:1989dd, Beneke:1997zp, Smirnov:1990rz}
is a convenient way to obtain the leading power, whereas it is getting
rather complicated for higher powers because of the many contributing
regions and because it is difficult to automatize. 
Furthermore it is a very non-trivial task to make sure that one has
not forgotten or counted twice any region. However in the Euclidean
limit, where no collinear divergences arise,
automatizations exist, which rely on graph theory 
\cite{Chetyrkin:1996my,Seidensticker:1999bb}.
Another way to expand Feynman integrals, which has been proposed and 
worked out in 
\cite{Remiddi:1997ny, Pilipp:2007wm,Pilipp:2007mg,Boughezal:2007ny},
is based on differential equations.
Differential equation
techniques, which has been proposed first in \cite{Kotikov:1990kg},
is easy to automatize in a computer algebra
system. This makes it a convenient method to obtain
subleading powers, whereas the leading power is in most cases needed
as an input like a boundary condition. Another limitation is the fact
that this method relies on a correct ansatz in terms of powers of the
expansion parameter. However it
is a priori not obvious which powers of the expansion parameter occur
(e.g.\ only integer powers or also half-integer powers). 

In the present paper I present a semi-numerical method, that provides
the power expansion of Feynman integrals by giving explicit
expressions of the expansion coefficients in form of finite integrals
the can be solved numerically. In particular this method gives the
contributing powers of the expansion parameter, from where one can read off
the correct ansatz to solve the differential equations that determine
the set of Feynman integrals. 

The algorithm that is worked out in the present paper combines sector
decomposition 
\cite{Binoth:2000ps,Heinrich:2002rc,Binoth:2003ak,Heinrich:2008si}
with Mellin-Barnes techniques. It is completely independent from
any power counting argument such that it can be used as a cross check for
method of regions. This is very useful in cases, where  
method of regions becomes involved because of many contributing regions.

The paper is organized as follows. In Section \ref{s1} 
the algorithm is explained in detail. In Section
\ref{s2} I apply this algorithm to a set of two Feynman
integrals, that are power expanded by differential equation
techniques, where the leading powers are obtained by method of
regions. I will show explicitly how this algorithms gives the
correct ansatz for the differential equations and provides a
non-trivial check for method of regions.

\section{Algorithm \label{s1}}
We follow the steps of Section 2 of \cite{Binoth:2000ps}. We start
with a $L$-Loop Feynman integral
\begin{equation}
G=\int\prod_{i=1}^L\frac{d^Dk_i}{(2\pi)^D}\frac{1}{P_1\ldots P_N}
\end{equation}
which using the Feynman
parameterization
\begin{equation}
\frac{1}{P_1\ldots P_N}=\Gamma(N)\int_0^1d^Nx\,
\frac{\delta\left(1-\sum_{n=1}^N x_n\right)}
{\left(
x_1P_1+\ldots +x_NP_N\right)^N}
\end{equation}
can be cast into the form:
\begin{equation}
G=\Gamma(N)\int d^Nx\,\delta(1-\sum_{n=1}^N x_n)
\int\prod_{i=1}^L\frac{d^D k_i}{(2\pi)^D}
\left[\sum_{j,l=1}^{L} k_j\cdot k_l M_{jl}-2
\sum_{j=1}^{L}k_j\cdot Q_j+J\right]^{-N}.
\label{1.1}
\end{equation}
We define $D=4-2\epsilon$ as usual.
After performing the integration over the loop momenta we obtain:
\begin{equation}
G=(-1)^N\left(\frac{i}{(4\pi)^{D/2}}\right)^L
\Gamma(N-LD/2)\int d^Nx\,\delta(1-\sum_{n=1}^N x_n)
\frac{U^{N-(L+1)D/2}}{F^{N-LD/2}},
\label{1.2}
\end{equation}
where
\begin{equation}
F=-\det(M)\left[
J-\sum_{j,l=1}^{L}Q_j\cdot Q_l M_{jl}^{-1}\right]
\label{1.3}
\end{equation}
and 
\begin{equation}
U=\det(M).
\end{equation}
Let us assume (\ref{1.3}) contains the parameter $\lambda$, in which
we want to expand (\ref{1.1}). Using the Mellin-Barnes representation
\cite{Smirnov:2002pj}
\begin{equation}
\frac{1}{(X_1+X_2)^x}=\frac{1}{\Gamma(x)}\frac{1}{2\pi i}
\int_{-i\infty}^{i\infty} ds\,\Gamma(-s)\Gamma(s+x)X_1^s X_2^{-s-x},
\end{equation}
where the integration contour over $s$ has to be chosen such that 
\begin{displaymath}
-x<\Re(s)<0,
\end{displaymath}
we modify (\ref{1.2}) in the following way
\begin{equation}
\begin{split}
G=&(-1)^N\left(\frac{i}{(4\pi)^{D/2}}\right)^L
\frac{1}{2\pi i}\int_{-i\infty}^{i\infty}ds\,
\lambda^s
\Gamma(-s)\Gamma(s+N-LD/2)\\
&\times\int d^Nx\,\delta(1-\sum_{n=1}^N x_n)
U^{N-(L+1)D/2}F_1^sF_2^{-s-N+LD/2},
\end{split}
\label{1.4}
\end{equation}
where
\begin{equation}
F = \lambda F_1+F_2.
\end{equation}
The main idea behind the procedure below is the following:
By closing the integration path to the right
hand side of the imaginary axis we sum up all the residua on the
positive real axis and obtain an expansion in $\lambda$. Powers of
$\ln\lambda$ appear because of poles of order higher than one and
because of terms of the form $\lambda^{A-B\epsilon}$ in the
expansion in $\lambda$. These terms turn after expanding in 
$\epsilon$ into powers of $\ln\lambda$.

We continue with part I and II of \cite{Binoth:2000ps}. First we split
the integral over the Feynman parameters into
\begin{equation}
\int d^Nx = \sum_{l=1}^N\int d^Nx \prod_{\stackrel{j=1}{j\ne l}}^N\theta(x_l-x_j)
\end{equation}
and integrate out the $\delta$-function by the substitution
\begin{equation}
x_j=\left\{ 
\begin{array}{l@{\quad}l}
x_l t_j & j<l\\
x_l     & j=l\\
x_l t_{j-1} & j>l
\end{array}
\right.
\label{1.4.1}
\end{equation}
such that we obtain
\begin{equation}
G=(-1)^N\left(\frac{i}{(4\pi)^{D/2}}\right)^L
\frac{1}{2\pi i}\int_{-i\infty}^{i\infty}ds\,
\lambda^s
\Gamma(-s)\Gamma(s+N-LD/2)
\sum_{l=1}^N \int_0^1 d^{N-1}t\,G_l,
\label{1.5}
\end{equation}
where 
\begin{equation}
G_l=U_l^{N-(L+1)D/2}F_{1,l}^sF_{2,l}^{-s-N+LD/2}
\end{equation}
is obtained by the substitution (\ref{1.4.1}).
In (\ref{1.5}) the integration over small $t$ leads to poles in $s$. 
This
behavior is made explicit, if we follow the steps of Part II of 
\cite{Binoth:2000ps}: Look
for a minimal set $\{t_{\alpha_1},\ldots,t_{\alpha_r}\}$ such that
$U_l$, $F_{1,l}$ or $F_{2,l}$ vanish, if these parameters are set
to zero. We decompose the integral into $r$ subsectors
\begin{equation}
\int_0^1 d^{N-1}t=\int_0^1 d^{N-1}t
\sum_{k=1}^r\prod_{\stackrel{j=1}{j\ne k}}^r \theta(t_{\alpha_k}-t_{\alpha_j})
\end{equation}
and substitute
\begin{equation}
t_{\alpha_j}\to\left\{
\begin{array}{l@{\quad}l}
t_{\alpha_k} t_{\alpha_j} & j\ne k\\
t_{\alpha_k}             & j = k
\end{array}\right.,
\end{equation}
which leads to the Jacobian factor $t_{\alpha_k}^{r-1}$.
Now we are able to factorize out $t_{\alpha_k}$ from $U_l$, $F_{1,l}$
or $F_{2,l}$. After repeating these steps, until $U_l$, $F_{1,l}$
and $F_{2,l}$ contain terms that are constant in $\vec{t}$, we end up 
with integrals over the Feynman parameters of the form
\begin{equation}
\sum_{l=1}^N\sum_k \int_0^1 d^{N-1}t\left(
\prod_{j=1}^{N-1}t_j^{A_j-B_j\epsilon-C_js}\right)
U_{lk}^{N-(L+1)D/2}F_{1,lk}^sF_{2,lk}^{-s-N+LD/2},
\label{1.6}
\end{equation}
where $U_{lk}$, $F_{1,lk}$ and $F_{2,lk}$ contain terms that are
constant in $\vec{t}$. The procedure above can in principal lead to
infinite loops. This problem was addressed in
\cite{Bogner:2007cr,Smirnov:2008py}, where algorithms are proposed
that avoid these endless loops by choosing appropriate subsectors. 
I have not yet faced any endless loop in the problems I
dealt with. However one should keep in mind that they can
occur and adapt the implementation of the algorithm if needed. 

From (\ref{1.6}) we can read off that the poles in $s$ are located at:
\begin{equation}
s_{jn}=\frac{1+n+A_j-B_j\epsilon}{C_j},
\label{1.7}
\end{equation}
where $n\in \mathbb{N}_0$. Eq.~(\ref{1.7}) becomes clear if one Taylor
expands in (\ref{1.6}) the terms outside the brackets with respect to 
$t_j$ and performs the integration.

In (\ref{1.5}) we have to choose the contour of the integration over
$s$ such that the integration over the Feynman parameters $t_j$
converges. This leads to the condition 
\begin{equation}
A_j-B_j\epsilon-C_j \Re(s)>-1\quad\forall j.
\label{1.7.1}
\end{equation}
The poles in (\ref{1.7}) that have to be taken into account are those that
are located on the right hand side of the integration contour, i.e.
\begin{equation}
\Re(s)<s_{jn}.
\label{1.7.2}
\end{equation} 
From (\ref{1.7}) and (\ref{1.7.1}) we conclude that (\ref{1.7.2})
is fulfilled if and only if $C_j>0$.

In the next step we calculate the residue of (\ref{1.6}) at
$s_{jn}$. We write the $k$'th Feynman integral in the form
\begin{equation}
\int_0^1 dt_k \,t_k^{A^\prime- B^\prime\epsilon-C^\prime(s-s_{jn})}
\mathcal{I}(t_k,s)
\label{1.8}
\end{equation}
and note that this term is singular in $s-s_{jn}$ if and only if
\begin{equation}
 B^\prime=0\quad\text{and}\quad A^\prime\le -1. 
\label{1.9}
\end{equation}
So following Part III of 
\cite{Binoth:2000ps} we expand $\mathcal{I}(t_k,s)$ around $t_k=0$
and obtain 
\begin{equation}
\mathcal{I}(t_k,s)=\sum_{p=0}^{-A^\prime-1} 
    \mathcal{I}^{(p)}(s)\frac{t_k^p}{p!}+R(t_k,s),
\label{1.10}
\end{equation}
with a rest term $R(t_k,s)=\mathcal{O}(t^{-A^\prime})$,
such that (\ref{1.8}) becomes
\begin{equation}
\sum_{p=0}^{-A^\prime-1}\frac{1}{A^\prime+1+p-C^\prime(s-s_{jn})}
\frac{\mathcal{I}^{(p)}(s)}{p!}+
\int_0^1 dt_k\,t_k^{A^\prime-C^\prime(s-s_{jn})}R(t_k,s),
\label{1.11}
\end{equation}
where we used that $B^\prime=0$.
We repeat this procedure for all $k$ where condition (\ref{1.9}) is
fulfilled. The remaining integrals do not diverge for $s=s_{jn}$. So
it is save to expand them around $s-s_{jn}$ and we can easily
calculate the residue at $s=s_{jn}$.

What is left is to calculate the Laurent expansion in
$\epsilon$. From the previous procedure we obtain terms of the form 
\begin{equation}
\int_0^1 d^n t\,\left(\prod t_j^{A_j^{\prime\prime}-B_j^{\prime\prime}\epsilon}
(\ln t_j)^{\alpha_j}\right) I(\vec{t},\epsilon)
\label{1.12}
\end{equation}
The logarithms $(\ln t_j)^{\alpha_j}$ arise from taking the residues of
terms of the form
$\frac{t_j^{-C^\prime(s-s_{jn})}}{(s-s_{jn})^m}$ with $m\ge 2$. 
In (\ref{1.12}) we wrote these logarithms
explicitly such that we can expand $I(\vec{t},\epsilon)$ around $t_j=0$.
The poles in $\epsilon$ in (\ref{1.12}) originate from integrals
\begin{equation}
\int_0^1 d t_j\,t_j^{A_j^{\prime\prime}-B_j^{\prime\prime}\epsilon}
(\ln t_j)^{\alpha_j} \mathcal{I}(t_j,\epsilon)
\label{1.13}
\end{equation}
with $A_j^{\prime\prime}\le -1$. Repeating the procedure above we expand
\begin{equation}
\mathcal{I}(t_j,\epsilon)=\sum_{p=0}^{-A_j^{\prime\prime}-1} 
    \mathcal{I}^{(p)}(\epsilon)\frac{t_j^p}{p!}+R(t_j,\epsilon)
\label{1.14}
\end{equation}
and obtain for (\ref{1.13})
\begin{equation}
\sum_{p=0}^{-A_j^{\prime\prime}-1}
\frac{(-1)^{\alpha_j}(\alpha_j+1)!}
{(1+p+A_j^{\prime\prime}-B_j^{\prime\prime}\epsilon)^{\alpha_j+1}}
\frac{\mathcal{I}^{(p)}(\epsilon)}{p!}
+\int_0^1 dt_j\, t_j^{A_j^{\prime\prime}-B_j^{\prime\prime}\epsilon}(\ln t_j)^{\alpha_j}
R(t_j,\epsilon).
\label{1.15}
\end{equation}
All the remaining integrals over $t_j$ are finite and can in principle
be calculated numerically. Finally the original integral $G$ in (\ref{1.1})
obtains the form
\begin{equation}
G = \sum_{i,m,n} \epsilon^i \lambda^{m}(\ln\lambda)^n I_{i,m,n},
\label{1.16}
\end{equation}
where the $I_{i,m,n}$ contain finite integrals that can be numerically
evaluated. The logarithms $(\ln\lambda)^n$ arise both due to poles of
higher order in the Mellin-Barnes parameter and to the expansion in
$\epsilon$ from terms of the form
$\lambda^\epsilon/\epsilon^n$. Depending on the values of $C_j$ in
(\ref{1.7}) the sum over $m$ does not only run over integer numbers but
also over numbers of the form 
$$\frac{1+n+A_j}{C_j},$$ 
where $n$ is integer. I stress that even if a
numerical evaluation of the integrals $I_{i,m,n}$ is not possible, we
can obtain non-trivial statements about the power expansion of $G$ from 
(\ref{1.16}) together with (\ref{1.7}). That is to say (\ref{1.7})
gives us information about the possible powers of $\lambda$ e.g.\ we
know if we only get integer powers or also powers of
$\sqrt{\lambda}$. And from (\ref{1.16}) we can read off up to which power
$\ln\lambda$ appears. As we will see in the next section this
information will prove to be useful to obtain the power expansion by
means of differential equations.

\section{Example: Power expansion of Feynman integrals by 
differential equation techniques
\label{s2}}
\begin{figure}
\begin{center}
\resizebox{0.5\textwidth}{!}{\includegraphics{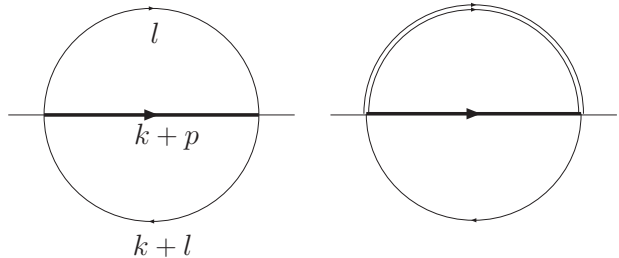}}
\end{center}
\caption{Sunrise diagrams. The thick line denotes a propagator of mass
  $M$, while the thin lines stand for mass $m$. The double line
  denotes that the propagator is to be taken squared.}
\label{f1}
\end{figure}
The idea to get the expansion of Feynman integrals by differential
equations has been proposed and worked out in \cite{Remiddi:1997ny, 
Pilipp:2007wm,Pilipp:2007mg,Boughezal:2007ny}.
By the following example we will see that the algorithm shown in the
last section will give us the correct ansatz to solve the given system
of differential equations and help us with the calculation of the
initial conditions.
We start with the integrals given by Fig.~\ref{f1}, where we assume
$p^2=M^2$:
\begin{eqnarray}
I_1&=&\int\frac{d^Dk}{(2\pi)^D}\frac{d^Dl}{(2\pi)^D}
\frac{1}{(k^2+2k\cdot p)\left((k+l)^2-m^2\right)
\left(l^2-m^2\right)}
\nonumber
\\
I_2&=&\int\frac{d^Dk}{(2\pi)^D}\frac{d^Dl}{(2\pi)^D}
\frac{1}{(k^2+2k\cdot p)\left((k+l)^2-m^2\right)
\left(l^2-m^2\right)^2}.
\label{2.1}
\end{eqnarray}
Let us assume that we want to expand these integrals in
$\lambda=m^2/M^2$ and need the result up to order $\mathcal{\epsilon}$. 
For simplicity let us also set
$M^2=1$ and $m^2=\lambda$. Using integration-by-parts identities
\cite{Chetyrkin:1981qh,Tkachov:1981wb,Laporta:2001dd}, we
get the following differential equations for $I_1$ and $I_2$:
\begin{eqnarray}
\frac{d}{d\lambda} I_1 &=& 
h_{11}I_1+h_{12}I_2+g_1
\nonumber\\
\frac{d}{d\lambda} I_2 &=& 
h_{21}I_1+h_{22}I_2+g_2
\label{2.2}
\end{eqnarray}
with
\begin{equation}
h = 
\left(
\begin{array}{cc}
0 & 2\\
\frac{1}{2\lambda(1-\lambda)}&\frac{1-3\lambda}{2\lambda(1-\lambda)}
\end{array}
\right)+
\epsilon
\left(
\begin{array}{cc}
0 & 0\\
-\frac{7}{4\lambda(1-\lambda)}&
\frac{-2+4\lambda}{\lambda(1-\lambda)}
\end{array}
\right)+
\epsilon^2
\left(
\begin{array}{cc}
0 & 0\\
0 &
\frac{3}{2\lambda(1-\lambda)}
\end{array}
\right)
\label{2.3}
\end{equation}
and
\begin{eqnarray}
g_1 &=& 0
\nonumber\\
g_2 &=& 
\frac{(1-\epsilon)^2}{4\lambda^2(1-\lambda)}
\bigg[
\int\frac{d^Dk}{(2\pi)^D}\frac{d^Dl}{(2\pi)^D}
\frac{1}{\left((k+l)^2-\lambda\right)\left(l^2-\lambda\right)}-
\nonumber\\
&&
\int\frac{d^Dk}{(2\pi)^D}\frac{d^Dl}{(2\pi)^D}
\frac{1}{(k^2+2k\cdot p)\left(l^2-\lambda\right)}
\bigg]
\nonumber\\
&=&\frac{1}{(4\pi)^D}\Gamma(\epsilon)^2
\frac{\lambda^{1-2\epsilon}-\lambda^{-\epsilon}}{4\lambda(1-\lambda)},
\label{2.4}
\end{eqnarray}
where (\ref{2.3}) and (\ref{2.4}) are exact in $\lambda$ and
$\epsilon$.
By defining 
\begin{eqnarray}
I_\alpha &=& \sum_{i,j,k} I_{\alpha,i}^{(j,k)} 
\epsilon^i \lambda^j (\ln\lambda)^k
\nonumber\\
h_{\alpha\beta}&=&\sum_{i,j} h_{\alpha\beta,i}^{(j)}
\epsilon^i\lambda^j
\nonumber\\
g_\alpha&=&\sum_{i,j,k} g_{\alpha,i}^{(j,k)}
\epsilon^i\lambda^j(\ln\lambda)^k
\label{2.5}
\end{eqnarray}
(\ref{2.2}) becomes
\begin{equation}
  0=
  (j+1)I_{\alpha,i}^{(j+1,k)}+(k+1)I_{\alpha,i}^{(j+1,k+1)}-
  \sum_{\beta=1,2}\sum_{i^\prime=0}^2\sum_{j^\prime=-1}^j
    h_{\alpha\beta,i^\prime}^{(j^\prime)}
    I_{\beta,i-i^\prime}^{(j-j^\prime,k)}
    -g_{\alpha,i}^{(j,k)}.
  \label{2.6}
\end{equation}
In (\ref{2.5}) we have not yet specified which values the summation index
$j$ takes and up to which maximum value the finite sum over $k$
runs. 
By implementing the steps of the last section, which led to
(\ref{1.7}), in a computer algebra system 
we obtain from (\ref{1.7}) that $I_1$ comes with the powers of $\lambda$
\begin{equation}
\lambda^n,\quad \lambda^{n+1-\epsilon},\quad 
\lambda^{\frac{n+3}{2}-2\epsilon}
\label{2.7}
\end{equation}
and $I_2$ with
\begin{equation}
\lambda^n,\quad \lambda^{n-\epsilon},\quad 
\lambda^{\frac{n+1}{2}-2\epsilon},
\label{2.8}
\end{equation}
where $n \in \mathbb{N}_0$. From (\ref{2.7}) and (\ref{2.8}) we read
off that $j$ takes the values $0, 1/2, 1, \ldots$. In (\ref{2.6})
integer-valued and half-integer-valued $j$ do not mix. So we would
have missed powers of $\sqrt{\lambda}$, if we had made the na\"{\i}ve
ansatz that $I_{1,2}$ only come with integer powers of
$\lambda$. Now one could argue that $\sqrt{\lambda}$ is already
contained in the sum over $\ln\lambda$. However in order to solve
(\ref{2.6}) we have to assume that there exists $k_\text{max}$ such
that $I^{(j,k)}_{\alpha,i}=0$ for all
$k > k_\text{max}$. A computer algebra analysis of the algorithm in
the previous section tells us that in our special case
$k_\text{max}=3$.

Solving (\ref{2.6}) up to $\mathcal{O}(\epsilon)$ we note that we need 
$I_{1,i}^{(0,0)}$ and $I_{2,i}^{(\frac{1}{2},0)}$ as initial
conditions, which can be obtained by method of regions 
\cite{Smirnov:2002pj,Gorishnii:1989dd,Beneke:1997zp,Smirnov:1990rz}.
In the case of $I_{1,i}^{(0,0)}$ we note that only the region
participates where both integration momenta are hard: 
\begin{equation}
k^\mu=\mathcal{O}(1)\quad\text{and}\quad
l^\mu=\mathcal{O}(1).
\end{equation}
In this region we obtain
\begin{equation}
\int\frac{d^Dk}{(2\pi)^D}\frac{d^Dl}{(2\pi)^D}
\frac{1}{(k^2+2k\cdot p)(k+l)^2 l^2}=
\frac{1}{(4\pi)^D}
\frac{\Gamma(-1+2\epsilon)
\Gamma(\epsilon)\Gamma(1-\epsilon)^2
\Gamma(3-4\epsilon)}
{\Gamma(2-2\epsilon)\Gamma(3-3\epsilon)},
\label{2.9}
\end{equation}
which is the leading power of $I_1$. For $I_{2,i}^{(\frac{1}{2},0)}$
we need the region where both $k$ and $l$ are soft, i.e.\
\begin{equation}
k^\mu=\mathcal{O}(\sqrt{\lambda})\quad\text{and}\quad
l^\mu=\mathcal{O}(\sqrt{\lambda}).
\end{equation}
This region starts participating at 
$\mathcal{O}(\sqrt{\lambda})$:
\begin{equation}
\begin{split}
&\int\frac{d^Dk}{(2\pi)^D}\frac{d^Dl}{(2\pi)^D}
\frac{1}{(2k\cdot p)\left((k+l)^2-\lambda\right) 
\left(l^2-\lambda\right)^2}=\\
&\quad\quad
\frac{-1}{(4\pi)^D}
\frac{2^{-2\epsilon}\pi\Gamma\left(\epsilon-\frac{1}{2}\right)
\Gamma\left(2\epsilon-\frac{1}{2}\right)}{\Gamma(\epsilon)}
\lambda^{\frac{1}{2}-2\epsilon}.
\end{split}
\label{2.10}
\end{equation} 
By comparing these results to (\ref{2.7}) and (\ref{2.8}) we note 
that (\ref{2.9})
and (\ref{2.10}) correspond to definite poles in the Mellin-Barnes
representation i.e.\ at $s=0$ and $s=1/2-2\epsilon$. 
By (\ref{1.7}) and (\ref{1.11}) we can calculate the
coefficients of $\lambda^0$ and $\lambda^{\frac{1}{2}-2\epsilon}$ in
the $\lambda$-expansion of $I_1$ and $I_2$ numerically. 
This is a non-trivial
test that we have not forgotten a contributing region, which is in
general a problem of method of regions.

We normalize our integrals by multiplication 
with $(\exp(\gamma_\text{E})/(4\pi))^{2\epsilon}$ and obtain from the
solution of (\ref{2.6}) the analytical expansion in $\lambda$ and
$\epsilon$:
\begin{gather}
\begin{split}
&I_1 = \\
&\quad\quad\frac{1}{(4\pi)^4}\bigg[
-\frac{1}{2\epsilon^2}-\frac{5}{4\epsilon}-\frac{11}{8}-
\frac{5\pi^2}{12}+\epsilon\left(\frac{55}{16}-\frac{25\pi^2}{24}-
\frac{11}{3}\zeta(3)\right)+
\\
&\quad\quad
\lambda\bigg(
-\frac{1}{\epsilon^2}+\frac{-3+2\ln\lambda}{\epsilon}
-5+\frac{\pi^2}{2}+6\ln\lambda-(\ln\lambda)^2+
\\
&\quad\quad\quad
\epsilon\left(-3+\frac{3\pi^2}{2}+\frac{26}{3}\zeta(3)+
\left(14+\frac{\pi^2}{3}\right)\ln\lambda-3(\ln\lambda)^2+
\frac{(\ln\lambda)^3}{3}
\right)
\bigg)+
\\
&\quad\quad
\lambda^{\frac{3}{2}}\epsilon\frac{-16\pi^2}{3}+
\mathcal{O}(\lambda^2)
\bigg]+\mathcal{O}(\epsilon^2)
\end{split}
\nonumber
\\
\begin{split}
&I_2 = \\
&\quad\quad\frac{1}{(4\pi)^4}\bigg[
-\frac{1}{2\epsilon^2}+\frac{-1+2\ln\lambda}{2\epsilon}
+\frac{1}{2}+\frac{\pi^2}{4}+2\ln\lambda-\frac{1}{2}(\ln\lambda)^2+
\\
&\quad\quad\quad
\epsilon\left(\frac{11}{2}+\frac{11\pi^2}{12}+
\frac{13}{3}\zeta(3)+\left(4+\frac{\pi^2}{6}\right)\ln\lambda-
(\ln\lambda)^2+\frac{1}{6}(\ln\lambda)^3
\right)+
\\
&\quad\quad
\lambda^{\frac{1}{2}}(-4\epsilon\pi^2)+
\\
&\quad\quad
\lambda\bigg(
-1-\frac{\pi^2}{3}+\ln\lambda-\frac{1}{2}(\ln\lambda)^2+
\\
&\quad\quad\quad
\epsilon\left(11+\frac{2\pi^2}{3}-4\zeta(3)-3\ln\lambda-
\frac{1}{2}(\ln\lambda)^2+
\frac{1}{2}(\ln\lambda)^3
\right)
\bigg)+
\\
&\quad\quad
\lambda^{\frac{3}{2}}\epsilon\frac{4\pi^2}{3}+
\mathcal{O}(\lambda^2)
\bigg]+\mathcal{O}(\epsilon^2).
\end{split}
\label{2.11}
\end{gather}

On the other hand our numeric method of Section \ref{s1} gives
\begin{equation}
\begin{split}
&I_1 = \\
&\quad\quad 10^{-4}\bigg[
-\frac{0.20}{\epsilon^2}+\frac{-0.50}{\epsilon}-
2.2-4.5\epsilon+
\\
&\quad\quad
\lambda\bigg(
-\frac{0.40}{\epsilon^2}+\frac{-1.2+0.80\ln\lambda}{\epsilon}
-0.026+2.4\ln\lambda-0.40(\ln\lambda)^2+
\\
&\quad\quad\quad
\epsilon\left(8.9+
6.9\ln\lambda-1.2(\ln\lambda)^2+
0.13\ln\lambda)^3
\right)
\bigg)-
\\
&\quad\quad
21.\epsilon\lambda^{\frac{3}{2}}+
\mathcal{O}(\lambda^2)
\bigg]+\mathcal{O}(\epsilon^2)
\end{split}
\nonumber
\end{equation}
\begin{equation}
\begin{split}
&I_2 = \\
&\quad\quad 10^{-4}\bigg[
-\frac{0.20}{\epsilon^2}+\frac{-0.20+0.40\ln\lambda}{\epsilon}
+1.2+0.80\ln\lambda-0.20(\ln\lambda)^2+
\\
&\quad\quad\quad
\epsilon\left(7.9+2.2\ln\lambda-
0.40(\ln\lambda)^2+0.067(\ln\lambda)^3
\right)-
\\
&\quad\quad
16.\epsilon\lambda^{\frac{1}{2}}+
\\
&\quad\quad
\lambda\bigg(
-1.7+0.40\ln\lambda-0.20(\ln\lambda)^2+
\\
&\quad\quad\quad
\epsilon\left(5.1-1.2\ln\lambda-
0.20(\ln\lambda)^2+
0.20(\ln\lambda)^3
\right)
\bigg)+
\\
&\quad\quad
5.3\epsilon\lambda^{\frac{3}{2}}+
\mathcal{O}(\lambda^2)
\bigg]+\mathcal{O}(\epsilon^2),
\end{split}
\end{equation}
which is consistent with (\ref{2.11}).

\section{Conclusions}
By combining sector decomposition with Mellin-Barnes techniques I
developed an algorithm for power expanding Feynman integrals, where
the coefficients in the expansion are given by finite integrals. Even
if these integrals cannot be evaluated numerically, we can read off,
which powers of the expansion parameter
contribute and up to which power the logarithms occur. This
non-trivial information provides the correct ansatz for solving the
set of differential equations that determine the Feynman integrals.

Another application of the presented algorithm is testing method of
regions numerically. We have seen that every region, that has a
unique scaling in the expansion parameter, corresponds to a definite power 
in the Mellin-Barnes expansion. So it can be tested separately.
For method of regions it is often an involved problem to make sure not
to have missed or counted twice any region. This algorithm provides a
test of method of regions that is independent of any power counting
argument.

\begin{acknowledgments}
I thank Guido Bell and Christoph Greub for helpful discussions and
comments on the manuscript.
The author is partially supported by the Swiss National Foundation as
well as EC-Contract MRTN-CT-2006-035482 (FLAVIAnet).
\end{acknowledgments}

\end{document}